\documentclass[twocolumn,showpacs,nofootinbib]{revtex4-1} 
\usepackage{epsfig,amssymb,amsmath,latexsym} 
\usepackage{pifont} 
\usepackage{hyperref} 
 
\usepackage{color} 
\definecolor{nicered}{rgb}{0.7,0.1,0.1} 
\definecolor{nicegreen}{rgb}{0.1,0.5,0.1} 
  
\newcommand\eV{\text{eV}}

\newcommand\Mpc{\text{Mpc}}

\newcommand{\be}{\begin{equation}} 
\newcommand{\ee}{\end{equation}} 
\newcommand{\bea}{\begin{eqnarray}} 
\newcommand{\eea}{\end{eqnarray}} 
 
\newcommand{\no}{\noindent} 
\newcommand{\nb}{\nonumber}

\newcommand{\de}{\partial}

\newcommand\m{\mu} 
\newcommand\n{\nu} 
\newcommand\g{\gamma} 
\renewcommand\d{\delta}

\newcommand\V{{\ensuremath{\cal V}}} 
\newcommand\U{{\ensuremath{\cal U}}} 
\newcommand\C{{\ensuremath{\cal C}}}

\renewcommand\L{\ensuremath{\mathcal L}} 
 
\newcommand\ba{\begin{array}} 
\newcommand\ea{\end{array}}

\newcommand{\plm}{M_{\text{pl}}^2}

\newcommand\SEC[1]{\medskip\noindent{\sl\bfseries #1}}
 
\pacs{04.50.Kd,04.20.Fy} 
 
\begin{document} 
 
\title{Weak Massive Gravity} 
 
\author{D. Comelli$^a$, F. Nesti$^b$ and   L. Pilo$^{b,c}$} 
\affiliation{ 
  $^a$INFN - Sezione di Ferrara,  I-35131 Ferrara, Italy\\ 
  $^b$Dipartimento di Fisica, Universit\`a di L'Aquila, I-67010 L'Aquila\\ 
  $^c$INFN, Laboratori Nazionali del Gran Sasso, I-67010 Assergi, Italy 
} 
\date{\small \today} 
 
\begin{abstract} 
  \no We find a new class of theories of massive gravity with five propagating degrees of freedom
  where only rotations are preserved.  Our results are based on a non-perturbative and
  background-independent Hamiltonian analysis. In these theories the weak field approximation is
  well behaved and the static gravitational potential is typically screened \`a la Yukawa at large
  distances, while at short distances no vDVZ discontinuity is found and there is no need to rely on
  nonlinear effects to pass the solar system tests.  The effective field theory analysis shows that
  the ultraviolet cutoff is $(m M_{Pl})^{1/2}\simeq$ $ 1/\mu \rm m$, the highest possible.  Thus,
  these theories can be studied in weak-field regime at all the phenomenologically interesting
  scales, and are candidates for a calculable large-distance modified gravity.
\end{abstract} 
 
 

\maketitle

\SEC{Introduction.}  The long standing quest for a healthy modification of gravity at large
distances has recently been the target of renewed interest.  Since the work of Fierz and
Pauli~\cite{Fierz:1939ix} (FP) it was realized that in generic Lorentz-invariant (LI) massive
deformations of gravity 6 DoF are present already at linearized level around Minkowski space, the
sixth mode being a ghost.  While a tuning can get rid of this mode at linear level, it reappears at
the nonlinear level or around non-flat backgrounds~\cite{BD}.  It is remarkable that only recently a
nonlinear completion was found which propagates five DoF also in general. This was proposed first up
to the fourth order in perturbation theory~\cite{Gabadadze:2011} and then to the full nonlinear
level in~\cite{GF,DGT}.
At linearized level this theory reduces to the Lorentz-invariant FP theory, and shares with it the
failure to reproduce the correct light bending, at odds with General Relativity (GR) and
observations. The issue does not disappear in the limit of vanishing graviton mass, the so called
vDVZ discontinuity~\cite{DIS}.  A way out was proposed by Vainshtein in~\cite{Vainshtein}, by
arguing that nonlinear effects restore the correct GR behavior at short distances from a source.
While this mechanism has been verified in a number of models~\cite{vain2}, the theory has to rely on
strong nonlinearities even at the macroscopic solar system scales, where the gravitational potential
is small, as discussed for generic LI theories in~\cite{Dvali:2006su}.

A different perspective is provided by more general theories of massive gravity where Lorentz
invariance in the gravitational sector is not imposed, though the weak equivalence principle still
holds.  At the quadratic level a number of such theories that are free from the vDVZ discontinuity
is on the market~\cite{Rubakov,dub,PRLus,gaba,diego}.  Moreover, in some of them the cutoff beyond
which the theory cannot be trusted or where nonlinearities outbreak, has been argued to be the
highest possible~\cite{Rubakov}.  Clearly, if there exists nonlinear completions of such theories,
they represent serious candidates for a consistent and calculable large distance modifications of
gravity.
 
In this Letter we uncover a full class of nonlinear theories implementing these ideas.  The analysis
is based on a recent work~\cite{Comelli:2012vz} stating the general conditions for the propagation
of five DoF in massive gravity, whose main result is reviewed in the next section.  In the following
section we provide the general solution to these conditions.  In the remaining part we describe an
explicit example providing massive gravity theories which are free from the vDVZ discontinuity at
short distances, lead to screened (Yukawa) gravitational potential at large distances, propagate
five massive DoF both at linear and nonlinear level, and are calculable in the weak-field limit at
phenomenologically interesting distances.
 
\SEC{Massive gravities with 5 DoF.}  The generic Lagrangian for massive gravity is obtained by
adding a non-derivative potential $V$ for the metric $g_{\m\n}$,
\be  
L= \plm \int d^3x \sqrt{g}\,\Big[ R - \, m^2 \, V (g) +{\cal L}_{matter}\Big] \, ,
\label{eq:act} 
\ee 
and its propagating DoF were analyzed in~\cite{Comelli:2012vz} using the Hamiltonian formalism, with
the standard ADM decomposition~\cite{ADM} in terms of lapse $N$, shifts $N^i$ and spatial metric
$\g_{ij}$
\vspace*{-1ex}
\be 
g_{\mu \nu} = \begin{pmatrix} - N^2 + N^i N^j \g_{ij} & \g_{ij}N^j \\ 
\g_{ij}N^j  & \g_{ij} \end{pmatrix}. 
\ee 
The potential $V$ is regarded as a function of these variables, and it is useful to define $\V
\equiv m^2 \, N \, \det(\gamma)^{1/2} \, V $.  We will also denote $N^A =(N, \,N^i)$ with
$A=0,1,2,3$ (note $N^A$ is not a Lorentz 4-vector), and with $F_{AB\ldots}$ the derivative of any
$F$ with respect to $N^A$, $N^B$, etc.

As it stands, the presence of $V$ in the action violates diffeomorphism invariance, and in principle
six modes can propagate, with a (sixth) mode being typically a ghost. The conditions on the
potential ${\cal V}(N^A,\g^{ij})$ to have 5 propagating DoF and unbroken rotations were found
in~\cite{Comelli:2012vz}, and they turned out to be quite simple:

\smallskip

\noindent
$\bullet$ First, the four secondary constraints have to determine only three out of four auxiliary
fields~\cite{DGT,Comelli:2012vz}, leaving a new constraint. For this, the hessian of the potential
with respect to lapse and shifts must have rank~3
\be 
\qquad \det \V_{AB}=0\,,    \qquad (\mathop{\rm rank} \V_{AB}=3) \, .
\label{eq:hess} 
\ee 
So, $\V_{AB}$ has only one null eigenvector $\chi^A$: $\V_{AB} \,
\chi^B = 0$.

\pagebreak[3]

\smallskip

\noindent
$\bullet$ Second, the tertiary constraint has to be free of the last auxiliary
field~\cite{Comelli:2012vz}, to complete the new constraint in phase
space. This amounts to the following additional condition
\be 
\,  \, \tilde \V_i  + 2  \, \xi^A  \xi^j 
\,\frac{\de   \tilde \V_A}{\de \gamma^{ij}}  =0  \, .
\label{eq:add} 
\ee 
Here we defined $\tilde \V = \gamma^{-1/2} \V$ and $\xi^A=\chi^A/\chi^0$.  We are free to normalize
to $\chi^0=1$.

It can be shown that the same eqs.~(\ref{eq:hess}), (\ref{eq:add}) hold also for explicitly time
dependent potentials~\cite{long}.  One can verify that the Lorentz-invariant massive gravity theory
of~\cite{DGT,GF} nontrivially satisfies both (\ref{eq:hess}) and (\ref{eq:add}).  However, we show
here that a wider class of solutions with phenomenologically interesting features exist when the
potential is not Lorentz invariant but only rotational invariant. To this aim, let us the discuss
the general solution to~(\ref{eq:hess}) and (\ref{eq:add}).

\SEC{General solution.} We start from eq.~(\ref{eq:hess}), by recalling that it is a `homogeneous
Monge-Ampere' equation, a well studied problem in mathematical physics.  In the very clear
work~\cite{fairlie}, the general implicit solution was found by changing variables from $N^A$ to the
eigenvector $\chi^A$. In these variables, one finds that the potential must be an homogeneous
function of $\chi^A$ of degree zero. Then the key role is played by a generic function $
\U(\xi^i,\g_{ij})$, defined as $\U=\chi^A\, \tilde \V_A$ and linked to $\tilde\V_0$ by a Legendre
transformation
\bea 
&&\tilde \V_0= \U- \U_i \;\xi^i\,,\qquad \tilde \V_i= \U_i\,. 
\label{eq:legendre} 
\eea 
Note that $\tilde \V_A=\partial \tilde \V/\partial N^A$ while $\U_i=\partial \U/\partial \xi^i$.
Any function $ \U$ will lead to a solution of the Monge-Ampere equation~(\ref{eq:hess}),
provided~(\ref{eq:legendre}) can be integrated to find $\V$. Integrability is ensured~\cite{fairlie}
by an important implicit relation between $N^i$, $N$ and $\xi^i$,\footnote{We note \emph{en passant}
  that this implicit change of variables between $N^i$ and $\xi^i$ trivializes to a good extent the
  complicated Hamiltonian analysis of~\cite{Comelli:2012vz}, and contain the shift transformations
  of~\cite{GF}. This points to a possible important physical role of the $\xi^i$ variables, which in
  ADM language are linked to a 'spatial over temporal' speed of the foliation.}  which includes a
new generic function $\L(\xi^i,\,\g^{ij})$:
\be 
N^i - N \;\xi^i= \;\U^{ij}\; \L_j\,, 
\label{eq:xi} 
\ee 
where $ \U^{ij}\equiv( \U_{ij})^{-1}$ (inverse of the hessian matrix) and
$\L_i\equiv \partial_{\xi^i} \L$. For any $\L$, one should formally invert~(\ref{eq:xi}) to
determine $\xi^i(N^A,\g^{ij})$, and then integrate (\ref{eq:legendre}) in $N^0$, $N^i$ to find the
potential.
As a result, the general solution of~(\ref{eq:hess}) is expressed in terms of two generic functions
$\U$ and $\L$ of $\xi^i$ and $\g^{ij}$.  Note that up to now the variables $\gamma_{ij}$ inside $
\U$ and $\L$ are spectators.
 
Equation (\ref{eq:add})\ brings $\g_{ij}$ into the game. When written in terms of $\U$, it
 reads
\be 
\frac{\partial  \U}{\partial \xi^i} +2 \,\xi^j\frac{\partial  \U}{\partial \gamma^{ij}}=0\,, 
\ee 
which can be now readily solved by stating that $\U$ is a scalar function of a particular
combination of $\xi^i$ and $\g^{ij}$:
\be 
 \U= \U(X^{ij})\,\quad {\rm with}\quad  X^{ij}=\big(\g^{ij}-\xi^i\,\xi^{j}\big)\,,
\ee 
The last step is to find the potential by integrating the relation~(\ref{eq:legendre}).  These
equations admit, as integration constant, a generic function $\C(\g^{ij})$ to be added to $\tilde\V$.

Summarizing, we have found a whole new class of solutions propagating five DoF, which are
parametrized by three generic functions $\U(X^{ij})$, $\L(\xi^i,\g^{ij})$ and $\C(\gamma^{ij})$.  In
view of unbroken rotations, $\U$ has to be a scalar function of $X^{ik}\d_{kj}$, i.e.\ of its three
invariant traces, and similarly for $\C\equiv\C(\g^{ik}\d_{kj})$.

In order to integrate~(\ref{eq:legendre}) and find a potential $\V$ explicitly, some concrete $\L$
has to be chosen first, to solve the implicit relation (\ref{eq:xi}) between $\xi^i$ and $N^A$.  We
remark that in general the resulting potential, function of $N$, $N^i$ and $\g_{ij}$, will break
Lorentz symmetry. The conditions for Lorentz invariance, leading to the potentials of~\cite{DGT,GF},
consist in complicated additional constraints that will not be discussed here.  Instead, interesting
new solutions can be found by many choices of $\L$. The general analysis will be reported in a
separate publication~\cite{long} --- here, it is enough to consider the case $\L=0$, which already
leads to interesting features and phenomenology.

\SEC{An example of explicit theories.} For $\L=0$, eq.~(\ref{eq:xi}) leads to the simple relation
$\xi^i=N^i/N$.  This gives $X^{ij}=\big(\gamma^{ij}-N^iN^{j'}/N^2\big)\equiv g^{ij}$, which
remarkably coincides with the spatial part of the inverse metric.  It is pleasing that also
equations~(\ref{eq:legendre}) can be integrated and one finds the following large family of explicit
potentials
\be  
V= \U(g^{ik}\delta_{kj}) + N^{-1} \, \C(\gamma^{ik}\delta_{kj})\,. 
\label{eq:pot} 
\ee 

It is interesting that in this example, a closed form for the Hamiltonian of the system upon using
the secondary constraints can be found. Using the expressions in~\cite{Comelli:2012vz}, we find that
$\U$ cancels out and (suppressing boundary terms)
\be
H=M_P^2 \,m^2\int\!{\rm d}^3x\, \sqrt{\g} \,\C\,.
\ee
As a result, by choosing $\C$ a positive function, the Hamiltonian is positive definite.

\medskip 

A basic test of the above theories consists in analyzing the weak field expansion around flat space,
to see whether it is healthy and the phenomenological implications are consistent with the weak
field tests of gravity.
 
The conditions on the potential~(\ref{eq:pot}), to admit Flat Minkowski background $\eta_{\m\n}$ are
\be \label{flat}
\U\,|_{g=\eta}  =0,\quad \left[ 
\U'+ \C'-\frac{\C}{2} \right]_ {| g=\eta} = 0 \, ,
\ee 
where $|_{g=\eta}\equiv\{N\to1,N^i \to 0,\g_{ij}\to \delta_{ij}\}$ and $\de_{\gamma^{ij}} {\cal C}
\equiv {\cal C}' \delta_{ij}$, $\de_{\gamma^{ij}} {\U} \equiv {\U}' \delta_{ij}$ on Minkowski.
 
At quadratic level, the expansion of the potential leads to mass terms for the metric fluctuations
$h_{\m\n}=g_{\m\n}-\eta_{\m\n}$, which can be parametrized as
\be 
\frac{\plm}{2} 
\Big(m_0^2\,h_{00}^2+2\,m_1^2\,h_{0i}^2-m_2^2\,h_{ij}^2+m_3^2\,h_{}^2-2\,m_4^2\,h_{00}\,h\Big) 
\label{Lm} 
\ee 
with $h=h_{ii}$ and repeated spatial indices are summed. The quadratic expansion reflects the
breaking of Lorentz invariance in the full potential and gives the most general rotationally
invariant mass terms for the graviton.\footnote{The Lorentz-Invariant FP case is recovered for
  $m_0=0$, $m_i=m$.} They were studied in~\cite{Rubakov,dub}, where it was shown that absence of
ghosts and stability of the linearized theory can be achieved for $m_0=0$, together with the
conditions $m^2_2>0$ and $m^2_1>m^2_4>0$.  It is thus crucial to verify whether the
potential~(\ref{eq:pot}) can satisfy these criteria.  Expanding the potential~(\ref{eq:pot}), we get
the mass spectrum
 \bea \nonumber
 &&m_0^2=0\,, \qquad\qquad\qquad\quad\ \, \left. m_2^2 ={\cal O}_{ij,ij} \right |_{g=\eta} \\[.5ex]
&&m_1^2 = 2 \,m_4^2= \left.3\,\U'\right |_{g=\eta}\,,\quad  \left. m_3^2 =- {\cal O}_{ii,jj} \right |_{g=\eta}\quad \\[1ex]
\nonumber&& 
 {\cal O}_{i_1j_1,i_2j_2}\equiv \partial_{\g_{i_2j_2}} 
\left( \partial_{\g_{i_1j_1}}\V 
-\frac{1}{2}\;\g_{i_1j_1} \;\V \right)  \, .
\eea  
We thus find a strong correlation between $m_1$ and $m_4$, and their ratio $m_1^2/m_4^2=2$ is nicely
consistent with the stability conditions mentioned above.\footnote{Note, this relation is modified
  in the general case $\L\neq0$, and the stability conditions turn into conditions on
  $\L$~\cite{long}.}  Concerning the conditions $m_{2.4}^2>0$, the class of potentials is large and
they can generically be satisfied.  

As expected from the non-perturbative canonical analysis, conditions (\ref{eq:hess}) and
(\ref{eq:add}) to have 5 DoFs have led to $m_0^2=0$, which implies the absence of the ghost sixth
mode at the linear level. Needless to say, after the analysis of~\cite{Comelli:2012vz}, in these
theories five DoF propagate both at linear and at nonlinear level.  Notice the important role of the
function $\C$, i.e.\ the integration constant of (\ref{eq:legendre}): if $\C =0$ then $m_1^2 =0$ and
only 3 DoF propagate at the quadratic level, i.e. the theory is strongly coupled at all scales.
   
\smallskip

Still at quadratic level, stability of LB theories was studied also in FRW curved spacetime
in~\cite{diego}. For instance in deSitter space with curvature ${\rm H}^2$, five stable modes
propagate provided ${\rm H}^2< m_4^4/(3(m_4^2-m_3^2)+m_2^2)$.  This means that the graviton mass
scale has to be larger than the Universe curvature, or in other words that it can take physically
interesting values.

\SEC{St\"uckelberg Point of View.}  In any theory with a broken gauge symmetry, the symmetry can be
recovered by introducing extra St\"uckelberg fields.  In our case, in the unitary gauge,
the potential~(\ref{eq:pot}) depends on the rotationally invariant scalars made out of the 3D
matrices
%
$X^i_j\equiv g^{ik}\delta_{kj}$, $\Gamma^i_j\equiv\g^{ik}\delta_{kj}$,
%
as well as on the 3D scalar $N$.  The breaking of diffeomorphisms can be understood as due to the
explicit presence of $N$ and of the external `frozen' metric $\delta_{ij}$ in the potential, and can
be restored by using four Stuckelberg fields $\Phi$ and $\Phi^i$, for instance by promoting $X$,
$\Gamma$ and $N$ to manifestly gauge invariant quantities
(see~\cite{dub,Rubakov} for a thorough analysis).

In the standard expansion around Minkowski space, $\Phi= t + \phi^0$, $\Phi^i=x^i + \phi^i$, the
metric fluctuations $h_{\mu\nu}$ in the quadratic Lagrangian (\ref{Lm}) are replaced by the gauge
invariant combinations $H_{\m\n}=h_{\m\n} - 2 \de_{(\m} \phi_{\n)}$.  Because $m_0^2$ vanishes,
$\phi^0$ does not propagate. On the contrary the three spatial St\"uckelberg fields do.  One can
also decompose them in transverse and longitudinal parts, $\phi_i=A_i+\de_i\varphi$, which makes
clear the difference with the known Lorentz-invariant case.  Indeed, while in the Lorentz-invariant
case the longitudinal field $\varphi$ is made canonical through rescaling by $1/m^2M_p$, which leads
to low strong coupling scale discontinuity and Vainshtein radius, here all three fields are made
canonical by the minimal normalization $\sim 1/m\,M_p$.  In particular for $\varphi$ this fact is
due to the difference $m_1^2-m_4^2$ being nonzero (see~(\ref{eq:prop}) below) and is thus a strict
consequence of Lorentz-breaking in the gravitational sector.

Together with the standard canonization of the graviton this normalization leads to three important
consequences in the full theory: i) the cutoff of the theory is $\Lambda_2= (m \,
M_{pl})^{1/2}$~\cite{Rubakov}; ii) via mixing with the metric, the coupling of the St\"uckelberg
fields to matter sources vanishes as $m$, and thus no discontinuity is present; iii) again in the
presence of sources, there is no scale of nonlinearity (Vainshtein) larger than the standard
Schwarzschild radius $R_s$ (provided $\Lambda_2>1/R_s$).

It is also instructive to show the quadratic Lagrangian of canonical propagating fields in the
`decoupling limit', defined as $m\to 0$, $M_{Pl}\to\infty$ keeping $\Lambda_2\equiv\sqrt{m\,M_{Pl}}$
and $T/M_{Pl}$ finite. One has symbolically
\bea 
\nonumber {\cal L}_{eff}&=&h_c \Box\, h_c+
(\dot\varphi)^2-\frac{m_1^2(m_2^2-m_3^3)}{2 m_4^2 (m_1^2-m_4^2)}(\nabla\varphi)^2+\\[1ex]
&&{}+(\dot A^c_i)^2-\frac{m_2^2}{m_1^2}(\nabla_j A^c_i)^2+\frac{\hat{h}_c}{M_{Pl}}T 
\label{eq:prop}
\eea
which has two transverse tensor DoF in $h_c$, two in the transverse vector $A_i^c$ and one in the
scalar $\varphi_c$.  All these modes are massive, with different masses (they is not visible in the
decoupling limit, but see~\cite{Rubakov}).  Note that the scalar kinetic term in~(\ref{eq:prop}) is
singular in the LI limit, where all $m_i\to m$.  Lorentz-breaking is present in the dispersion
relations of vector and scalar, which can propagate at speeds different from the one of the light.
The presence of these three extra polarizations, in addition to the interesting massive nature of
transverse gravitational waves (see e.g.~\cite{Gumrukcuoglu:2012wt}), could thus lead to striking
signals already in the forthcoming gravitational waves searches.

\SEC{Phenomenology.}  Let us focus on the gravitatiaonal potential generated by a static source,
with a non-zero energy-momentum-tensor component $t_{00}$. 

At linearized level, the gauge invariant gravitational potential $\Phi=h_{00}/2$ is (we set $M_P=1$
here) \vspace*{-1ex}
\bea \nonumber
\Phi&=&\frac{\Delta+m_2^2\,\frac{m_2^2-3\,
    m_3^2}{m_2^2-m_3^2}}{2\,\Delta^2+
\Delta \,m_4^2\,\frac{m_4^2-4\,m_2^2}{m_2^2-m_3^2}+3\,m_4^4\, \frac{m_2^2}{m_2^2-m_3^2}}\;\,t_{00}\\ 
&\equiv& \left[\frac{A_1}{\Delta-M_1^2}+\frac{A_2}{\Delta-M_2^2}\right] \,\frac{t_{00}}{2},
\eea 
where the squared masses $M_{1,2}^2\sim m^2$ and the dimensionless coefficients $A_{1,2}$ in the
formal decomposition of the second line depend on the mass parameters $m_i^2$, and
$A_1+A_2=1$~\cite{Berezhiani:2009kv}.  The squared masses $M_{1,2}^2$ may turn out to be real or
complex.  As a result, the radial profile of the gravitational potential from a point source can be
Yukawa, oscillatory, damped (or even plain exponential or Newton, for some special choices of the
$m_i^2$.)  Once the squared masses $M_1^2$ and $M_2^2$ are real positive, the linearized potential
is a sum of two Yukawa terms:
\be
\label{eq:yuk}
\Phi=\frac{M}{r}\left( A_1\,e^{-M_1r}+A_2\,e^{-M_2r}\right).
\ee 
At small $r$ both gauge invariant potentials~\cite{Rubakov} reduce to the linearized GR result,
i.e.\ there is no discontinuity.

In the full nonlinear theories that we propose, we can go one step further. In fact, the crucial
difference with the Lorentz invariant case is that here we can use perturbation theory. This is
actually required to confront with standard post-Newtonian (PN) tests.  Using second order
perturbation theory, we find that the vacuum spherically symmetric solution can be written, in
appropriate Newtonian coordinates ($\tau$, $\rho$), in the form
\bea
\label{eq:pn}
ds^2 = 
&&
{}-\!\left [1 -\!\frac{2M}{\rho}+m^2\rho^2\Bigg(c_1\frac M\rho + c_2 \frac{M^2}{\rho^2}\ln(m\rho)\Bigg)
\right] d \tau^2\   
\nb 
\\[1ex]
 &&\!\!\!\!\!\!\!\!\!\!\!\!\!\!\!
{}\!+\rho^2 d\Omega^2 
+\!\left[1 -\!\frac{2 M}{\rho}+m^2\rho^2\Bigg(c_3\frac M\rho + c_4\frac{M^2}{\rho^2}\Bigg) \right]^{-1}\!\!\!\!\!\!
d\rho^2
\eea
where for convenience the metric has also been expanded to leading order in $m \rho \ll1$.
The constants $c_{i}$ depend on the chosen potentials $\U$ and $\C$.  This result shows that the
perturbation expansion near a source has no Vains\-htein cutoff, is valid as in GR down to $\rho\sim 2
M$, and is continuous in the $m\to0$ limit, also at nonlinear level.

The choice of Newtonian coordinates is allowed because the coupling with matter preserves general
covariance, and we also note that it enables us to show that the leading corrections start at
$O(m^2)$ (unlike what would appear from the expansion of~(\ref{eq:yuk}) at small $r$, where
$g_{\theta \theta}$ and $g_{\varphi \varphi}$ are not in the standard Newtonian
form~\cite{Rubakov}).


In the plain case~(\ref{eq:yuk}) of large-distance Yukawa falloff, a fairly strong bound on the
scale $m$ can be set from the existence of the largest gravitational bound
states~\cite{Goldhaber:1974wg}, which nowadays translates into the bound $1/m\gtrsim5\,\Mpc$, or
$m\lesssim10^{-30}\,\eV$.  This strong bound implies that the proposed models of massive gravity
automatically pass the solar system tests, both at Newtonian and at PN level, given the above
expressions~(\ref{eq:pn}).

As a phenomenological comment, we note that if $M_1$ and $M_2$ are not of the same scale, e.g.\ if
$M_1\gg M_2$, and $A_1<0$, then at intermediate distances the radial force is enhanced with the
respect to the Newtonian case, leading to the flattening of rotation curves also in the absence of
dark matter (before the Yukawa falloff at further larger distances).  In this case, PN tests may
become more relevant.  We leave this scenario for a future investigation.

Finally, let us briefly comment on cosmological solutions.  In LI massive gravity,
Friedmann-Robertson-Walker backgrounds simply do not exist~\cite{DAmico,cosm}.  Viable cosmologies
do exist in the bigravity approach~\cite{cosm,Volkov:2012cf}, where however cosmological
perturbations are strongly coupled~\cite{cosmpert}.  In the present models it is likely that some
extension will also be needed for cosmology (similar e.g.~to~\cite{D'Amico:2012zv}). Perturbations
may instead turn out to be well-behaved, thanks to the absence of strong nonlinearities.  We will
address these issues in a separate work~\cite{long}.

\smallskip
\newcommand\xx{\bfseries}

\SEC{Conclusions.}  We analyzed the general theories of massive gravity with five propagating DoF
and unbroken rotations, which can be candidates for healthy large distance modifications of gravity.
The conditions for these theories, formulated in~\cite{Comelli:2012vz}, were solved in general, in
terms of three free functions.  Then, as an example, we described a large class of models that can
be analyzed explicitly, and have very interesting features. Namely:
{\xx (a)}~Five DoF propagate both at linear and nonlinear level.
{\xx (b)}~Gravitational waves are massive, including the three new graviton polarizations.  These
propagate with Lorentz-breaking dispersion relations, and may lead to striking signals in
forthcoming gravitational wave experiments.
{\xx (c)}~In static solutions, at short distances ($m r\ll1$) the vDVZ discontinuity is absent,
and there is no need to rely on nonlinear effects to pass the solar system tests of gravity; at
large distances $m r\gg1$ the gravitational potential is typically screened \emph{\'a la} Yukawa,
which is the desired behaviour of massive gravity.
{\xx (d)}~In this case, one can set a limit on the graviton mass of $m\lesssim10^{-30}\,\eV$.
{\xx (e)}~As an effective theory the ultraviolet cutoff is $\Lambda_2=\sqrt{m\,M_{\text{pl}}}$, the
highest possible in the absence of a fundamental Higgs mechanism for gravity.  Because of the above
bound, this corresponds to distances of $\sim10^{-3}\,$mm, below which nonlinear and/or quantum
corrections may lead to deviations from GR. This range is of interest for short-distance gravity
probes~\cite{Adelberger:2009zz}.
{\xx (f)}~Aside from this cutoff, the weak field approximation near a macroscopic source of mass $M$
is valid down to the Schwarzschild radius $R_S=2M/M_{Pl}^2$, as in GR.  Thus, a big advantage with
respect to the Lorentz-invariant theories is that here modified gravity remains weak as GR in all
the phenomenologically interesting regimes.

Of course to seriously compete with GR, massive gravity has to do more than just pass the solar
system tests; it has to reproduce the many successes of GR, ranging from emission of gravitational
waves from binary systems to CMB temperature fluctuations. Having a calculable weakly coupled theory
at hand is of great help, if not a prerequisite, in carrying out the program.

Finally, we also remark that the conditions for 5 DoF, or $m_0^2=0$ at quadratic level, are not
stable under quantum corrections, if one is willing to consider them.  Violations of these
conditions spoil the counting of DoF, reintroducing the sixth mode, usually at the cutoff scale.
This issue is shared among most massive gravity theories, and calls for a proper symmetry protection
for the framework described here.


\begin{thebibliography}{99} 
 
 
\bibitem{Fierz:1939ix} 
  M.~Fierz and W.~Pauli, 
  \emph{Proc.\ Roy.\ Soc.\ Lond.}\  A {\bf 173}, 211 (1939). 
 
 
\bibitem{BD} 
  D.G.~Boulware and S.~Deser, 
  \emph{Phys.\ Lett.}\  B {\bf 40}, 227 (1972). 
 
\bibitem{Gabadadze:2011}   
C.~de Rham, G.~Gabadadze, A.J.~Tolley, 
  \emph{Phys.\ Rev.\ Lett.}\  {\bf 106}, 231101 (2011). 
 
\bibitem{GF} 
S.F.~Hassan and R.A.~Rosen, 
  \emph{Phys.\ Rev.\ Lett.}\  {\bf 108}, 041101 (2012);\;
  S.F.~Hassan, R.A.~Rosen and A.~Schmidt-May, 
  \emph{JHEP} {\bf 1202}, 026 (2012) ;
S.F.~Hassan and R.A.~Rosen, 
JHEP {\bf 1204} (2012) 123
 
 
\bibitem{DGT}  
  C.~de Rham, G.~Gabadadze and A.~Tolley, 
  arXiv:1107.3820 [hep-th]. 


\bibitem{DIS} 
H.~van Dam and M.J.G.~Veltman, 
\emph{Nucl.\ Phys.\ }  B {\bf 22} (1970) 397; 
Y.~Iwasaki, 
\emph{Phys.\ Rev.\ }  D {\bf 2} (1970) 2255; 
 V.I.Zakharov, \emph{JETP Lett.} {\bf 12} (1971) 198. 
 
 
\bibitem{Vainshtein} 
  A.I.~Vainshtein, 
  \emph{Phys.\ Lett.}\  B {\bf 39}, 393 (1972);
 
\bibitem{vain2} 
  E.~Babichev, C.~Deffayet and R.~Ziour, 
  \emph{Phys.\ Rev.\ Lett.}\  {\bf 103}, 201102 (2009); 
  N.~Kaloper, A.~Padilla and N.~Tanahashi, 
  \emph{JHEP} {\bf 1110}, 148 (2011);
G.~Chkareuli and D.~Pirtskhalava,
 Phys.\ Lett.\ B {\bf 713}, 99 (2012).
 

\bibitem{Dvali:2006su} G.~Dvali,
  New J.\ Phys.\  {\bf 8} (2006) 326
  [hep-th/0610013].


\bibitem{Rubakov} 
V.A.~Rubakov, 
arXiv:hep-th/0407104. 
 
 
\bibitem{dub} 
  S.L.~Dubovsky, 
\emph{JHEP} {\bf 0410}, 076 (2004); 
V.A.~Rubakov and P.G.~Tinyakov, 
  \emph{Phys.\ Usp.}\  {\bf 51}, 759 (2008). 
 
 

\bibitem{gaba}    
  G.~Gabadadze and L.~Grisa, 
  \emph{Phys.\ Lett.}\  B {\bf 617} (2005) 124; 
  L.~Grisa, 
  \emph{JHEP} {\bf 0811} (2008) 023. 
 

\bibitem{PRLus} 
  Z.~Berezhiani, D.~Comelli, F.~Nesti and L.~Pilo, 
  \emph{Phys.\ Rev.\ Lett.}\  {\bf 99}, 131101 (2007). 
 
 
\bibitem{diego} 
 D.~Blas, D.~Comelli, F.~Nesti, L. Pilo, 
 \emph{Phys.\ Rev.}\  {\bf D80}, 044025 (2009). 
 
\bibitem{Comelli:2012vz} 
D.~Comelli, M.~Crisostomi, F.~Nesti and L.~Pilo,
  Phys.\ Rev.\ D {\bf 86}, 101502 (2012).
 
   
 
\bibitem{ADM}  
  R.L.~Arnowitt, S.~Deser and C.W.~Misner, 
  gr-qc/0405109. 
 
 
 \bibitem{HGS} 
  N.~Arkani-Hamed, H.~Georgi and M.D.~Schwartz, 
  Annals Phys.\  {\bf 305}, 96 (2003). 

\bibitem{long}D.~Comelli, F.~Nesti, L. Pilo, to appear. 

\bibitem{fairlie} D.B. Fairlie and A.N. Leznov, \emph{Journal of Geometry and Physics} {\bf 16} 
  (1995) 385--390.  
 
\bibitem{Berezhiani:2009kv}  
  Z.~Berezhiani, F.~Nesti, L.~Pilo and N.~Rossi, 
  JHEP {\bf 0907}, 083 (2009) .

\bibitem{Goldhaber:1974wg}
  A.~S.~Goldhaber and M.~M.~Nieto,
  Phys.\ Rev.\ D {\bf 9} (1974) 1119.



\bibitem{Gumrukcuoglu:2012wt}
  A.~E.~Gumrukcuoglu, S.~Kuroyanagi, C.~Lin, S.~Mukohyama and N.~Tanahashi,
  Class.\ Quant.\ Grav.\  {\bf 29} (2012) 235026.


\bibitem{DAmico} 
  G.~D'Amico, C.~de Rham, S.~Dubovsky, G.~Gabadadze, D.~Pirtskhalava and A.~J.~Tolley,
  Phys.\ Rev.\ D {\bf 84}, 124046 (2011).

\bibitem{cosm} 
D.~Comelli, M.~Crisostomi, F.~Nesti and L.~Pilo
  \emph{JHEP} {\bf 1203}, 067 (2012) 
  [Erratum-ibid.\  {\bf 1206}, 020 (2012)]. 
M.~von Strauss, A.~Schmidt-May, J.~Enander, E.~Mortsell and S.F.~Hassan, 
  \emph{JCAP} {\bf 1203}, 042 (2012);  
 M.S.~Volkov, 
  \emph{JHEP} {\bf 1201}, 035 (2012). 

\bibitem{Volkov:2012cf}
  M.~S.~Volkov,
  Phys.\ Rev.\ D {\bf 86} (2012) 061502.


\bibitem{cosmpert}
D.~Comelli, M.~Crisostomi and L.~Pilo, 
  \emph{JHEP} {\bf 1206}, 085 (2012) .




\bibitem{D'Amico:2012zv} 
  G.~D'Amico, G.~Gabadadze, L.~Hui and D.~Pirtskhalava,
  arXiv:1206.4253 [hep-th].


\bibitem{Adelberger:2009zz}
  E.~G.~Adelberger, J.~H.~Gundlach, B.~R.~Heckel, S.~Hoedl and S.~Schlamminger,
  Prog.\ Part.\ Nucl.\ Phys.\  {\bf 62} (2009) 102.


\end{thebibliography}
\end{document}